\documentclass[twocolumn]{aastex631}

\shorttitle{SNR Decoder Ring}
\shortauthors{Polin et al.}

\begin{document}

\title{Using Anisotropies as a Forensic Tool for Decoding Supernova Remnants}

\correspondingauthor{Abigail Polin}
\email{abigail@caltech.edu}

\author[0000-0002-1633-6495]{Abigail Polin}
\affiliation{The Observatories of the Carnegie Institution for Science, 813 Santa Barbara St., Pasadena, CA 91101, USA}
\affiliation{TAPIR, Walter Burke Institute for Theoretical Physics, 350-17, Caltech, Pasadena, CA 91125, USA}

\author[0000-0001-7626-9629]{Paul Duffell}
\affiliation{Department of Physics and Astronomy, Purdue University, 525 Northwestern Avenue, West Lafayette, IN 47907, USA}

\author[0000-0002-0763-3885]{Dan Milisavljevic}
\affiliation{Department of Physics and Astronomy, Purdue University, 525 Northwestern Avenue, West Lafayette, IN 47907, USA}
\affiliation{Integrative Data Science Initiative, Purdue University, West Lafayette, IN 47907, USA}

\begin{abstract}

We present a method for analyzing supernova remnants (SNRs) by diagnosing the drivers responsible for structure at different angular scales. First, we perform a suite of hydrodynamic models of the Rayleigh-Taylor instability (RTI) as a supernova collides with its surrounding medium. Using these models we demonstrate how power spectral analysis can be used to attribute which scales in a SNR are driven by RTI and which must be caused by intrinsic asymmetries in the initial explosion.  We predict the power spectrum of turbulence driven by RTI and identify a dominant angular mode which represents the largest scale that efficiently grows via RTI.  We find that this dominant mode relates to the density scale height in the ejecta, and therefore reveals the density profile of the SN ejecta.  If there is significant structure in a SNR on angular scales larger than this mode, then it is likely caused by anisotropies in the explosion.  Structure on angular scales smaller than the dominant mode exhibits a steep scaling with wavenumber, possibly too steep to be consistent with a turbulent cascade, and therefore might be determined by the saturation of RTI at different length scales (although systematic 3D studies are needed to investigate this).  We also demonstrate, consistent with previous studies, that this power spectrum is independent of the magnitude and length scales of perturbations in the surrounding medium and therefore this diagnostic is unaffected by ``clumpiness" in the CSM.

\end{abstract}


\section{Introduction}
\label{sec:intro}

Supernova remnants (SNRs) provide a powerful window through which to study the explosive deaths of stars. By taking careful measurements of the remnant, and combining resolved positions of features with their measured velocities from spectra, it is possible to map out the full 3D structure.  Combining images from several epochs makes it possible to see the evolution of this 3D structure over time. This process has been used successfully to develop 3D maps of several nearby SNRs (e.g. Cassiopia A \citep{2010ApJ...725.2038D,2015Sci...347..526M}, 1E 0102.2-7219 \citep{2010ApJ...721..597V,2018NatAs...2..465V}, N132D \citep{2011Ap&SS.331..521V,2020ApJ...894...73L}, and the Crab Nebula \citep{2010AJ....139.2083C,2021MNRAS.502.1864M}).  These observations have shown that SNRs exhibit turbulence, asymmetries and filaments that trace structure at a wide range of scales.  The physical mechanism (or combination of mechanisms) responsible for driving this detailed structure is not always clear. Possibilities include the turbulent flow driven by the Rayleigh-Taylor instability (RTI) \citep{1978ApJ...219..994C, 2008ARA&A..46..127H}, filaments carved out by large-scale magnetic fields \citep{2004A&A...423..253B}, and inhomogeneities traceable back to asymmetries in the initial explosion (such as asymmetric explosions \citep{2013A&A...552A.126W,2021A&A...645A..66O}, $^{56}$Ni bubbles \citep{Blondin2001,2021MNRAS.502.3264G} or jets \citep{1999ApJ...524L.107K,Bear2017}).  It is also debated what, if any, scales can be driven by variations, or clumps, in the gas surrounding the SN \citep{Celli2019}. 

The ``character" of the anisotropy is different between remnants; for example, the Tycho SNR exhibits a very spherical flow, with structure only on small angular scales.  Other remnants, like Cassiopeia A, have features on both large and small scales.  Generally these large and small-scale anisotropies have been discussed \citep{Fesen2001,Warren2005,2013ApJ...772..134M, 2021MNRAS.502.1694B,2021MNRAS.502.1864M, 2021MNRAS.504.2133N}, but differences in the character of anisotropy between different SNRs have not been quantified in a standard way.

The upcoming era of JWST observations provides opportunities to enhance the quality of SNR observations, at resolutions heretofore unprecedented at near- and mid-infrared wavelengths. These observations will likely reveal new, small scale behavior, and additional physics revealed only through infrared wavelengths sensitive to cool, unshocked ejecta \citep{Laming2020}. Interpreting this data in a manner which we can directly compare to theory and reveal the physics of the underlying explosions is vital.

We seek to develop a new way of characterizing the structure of SNRs by using power spectral analysis to diagnose sources of turbulence at different scales. We perform 2D hydrodynamics calculations of the evolution of SNRs and follow up with power spectral analysis to determine the power spectrum of fluid turbulence arising from the Rayleigh-Taylor Instability (RTI). In this manner we can demonstrate which portions of the SNR are dominated by turbulent effects due to instability, and which must arise from elsewhere, such as the initial asymmetry/anisotropy of the ejecta and/or CSM.

Section~\ref{sec:methods} describes our numerical methods and hydrodynamic models, and Section~\ref{sec:analysis} explains how we perform our power spectral analysis on the fluid ejecta properties.  Our results are presented in Section~\ref{sec:results}, and Section~\ref{sec:clumpy} discusses the effect of introducing clumping in the surrounding medium. The implications of our results for understanding SNR data is discussed in Section~\ref{sec:discussion}.

\section{Methods}
\label{sec:methods}

\subsection{Numerical Methods}
\label{subsec:jet}

Hydrodynamical calculations are performed using the JET code.  JET employs a moving-mesh hydrodynamical technique on a polar grid, with shearing radial tracks.  Each radial track behaves somewhat like a 1D Lagrangian hydrodynamics code, with neighboring tracks coupled by transverse fluid fluxes. The numerical method is based on the TESS code \citep{2011ApJS..197...15D} but is a distinct scheme that has been employed in many calculations in gamma ray bursts and supernovae (e.g. \cite{ 2013ApJ...775...87D}).

We evolve the non-relativistic hydrodynamical equations in conservation-law form:

\begin{equation}
    \partial_t \rho + \nabla \cdot ( \rho v ) = 0
\end{equation}
\begin{equation}
    \partial_t ( \rho \vec v ) + \nabla \cdot ( \rho v \vec v ) + \vec \nabla P = 0
\end{equation}
\begin{equation}
    \partial_t ( \frac12 \rho v^2 + \epsilon ) + \nabla \cdot ( (\frac12 \rho v^2 + \epsilon + P)v ) = 0
\end{equation}
where $\rho$ is mass density, $v$ is velocity, $P$ is pressure, and $\epsilon$ is the internal energy density.  The system is closed by an equation of state $\epsilon = 4P$ relevant for radiation-dominated flows.

The moving mesh reduces numerical diffusion and preserves contact discontinuities to high precision, making it ideal for studying Rayleigh-Taylor instability.  The JET code has already been employed to study RTI in a number of contexts \citep{2013ApJ...775...87D, 2014ApJ...791L...1D, 2016ApJ...821...76D, 2017ApJ...842...18D}.

\subsection{Chevalier Self Similar Solutions}
\label{sec:chevalier}

\cite{Chevalier1982SelfSimilar} found a self-similar solution that is commonly used to model the early phases of SNRs. The initial conditions describe the steeply varying outer layers of an ejecta colliding with a circumstellar medium (CSM):

\begin{equation}
    \rho_{\rm ejecta} \propto (r/t)^{-n} t^{-3}, ~~\rho_{\rm CSM} \propto r^{-s}.
\end{equation}

The ejecta velocity is assumed to be homologous ($v = r/t$) and collides with the stationary CSM. Where the ejecta collides with the CSM two shocks are formed, a forward shock that moves into the CSM and a reverse shock that moves into the ejecta. Between these two shock fronts the contact discontinuity becomes Rayleigh-Taylor unstable \citep{1978ApJ...219..994C}. These models most directly pertain to a young SNR, one where a reverse shock is present and has not yet propagated through the full SN ejecta.

Self-similar solutions for the turbulence can be found for a range of values of $n$ and $s$. A common choice is $n=7$, $s=2$ describing the steep outer layers of the ejecta with $\rho \propto r^{-7}$ colliding with a circumstellar wind with $\rho \propto r^{-2}$.  The choice $n=7$ appears to be consistent with some CSM interaction supernovae \citep{Chomiuk_2015}. 

One convenient advantage to analyzing a self-similar solution is that we can effectively run our simulation until the turbulence has reached a ``steady-state". This way we can perform our analysis on an ejecta that has reached a fully turbulent, statistically, self-similar state.

We use JET to perform a suite of Chevalier simulations for $s=0$ (a constant density environment) and $s=2$ (a wind environment) for $n=6-11$. We allow our ejecta to expand 6 orders of magnitude in time and the bulk of our analysis is performed on models with an initial resolution of $2048\times2048$ ($r\times\theta$), although JET may create more radial zones as needed. Section~\ref{sec:results} discusses the consequences of resolution in more detail. 

\section{Power Spectral Analysis}
\label{sec:analysis}

We propose analyzing SNR anisotropies by taking advantage of power spectral analysis techniques. Angular variations of physical qualities in the remnant will generically be represented by some function $f(\theta,\phi)$ which can be expressed as a linear combination of spherical harmonics:

\begin{equation}
    f(\theta,\phi) = \sum_{l,m}  a_{l m} Y_{l m}(\theta,\phi)
\end{equation}

We can calculate the amplitude associated with these variations by integrating:

\begin{equation}
    a_{l m} = \int f(\theta,\phi) Y^{*}_{l m}(\theta,\phi) d\Omega
\end{equation}

Now, while remnants are 3-dimensional the simulations in this work are 2D, so we can do an analogous analysis to understand the angular variations of physical quantities in the ejecta, $f(\theta)$, by using Legendre polynomials:

\begin{equation}
    f(\theta) = \sum_{l}  a_{l} P_l(cos \theta)
\end{equation}

We can understand the proper normalization for our amplitudes by standardizing $P_l(1) =1$ and recalling that Legendre polynomials are also orthogonal, so:

\begin{equation}
    \int_{-1} ^1 P_l(x) P_n(x) dx = \frac{2}{2l+1}\delta_{ln}
\end{equation}

Therefore the amplitude associated with $f(\theta)$ is:

\begin{equation}
    a_l = \frac{2l+1}{2} \int f(\theta) P_l (\cos\theta)d\cos\theta
\end{equation}

The power associated with the angular scale $\Delta\theta \sim \pi/l$ can then be measured by:

\begin{equation}
    C_l = \frac{1}{2l+1}|a_l|^2 .
\end{equation}

In 3D, the analogous expression would be

\begin{equation}
    C_l = \frac{1}{2l+1} \sum_{m = -l}^l |a_{lm}|^2
\end{equation}

In this work we examine the amplitudes and power associated with variations in both velocity and density. We do so by examining variations in angular scales by first pressure weighting radial quantities such that:

\begin{equation}
    \left< v(\theta) \right> = \frac{\int v(r, \theta)P(r, \theta)dr}{\int P(r, \theta)dr}
\end{equation}
and
\begin{equation}
    \left< \rho(\theta) \right> = \frac{\int \rho(r, \theta) P(r, \theta)dr}{\int P(r, \theta)dr}
\end{equation}

This pressure weighting helps to focus on anisotropies present in the shocked region.  Elsewhere the flow is cold and the pressure is nearly zero. We also normalize these quantities by the angle averaged value of velocity or density where appropriate.

\begin{figure}
    \centering
    \vspace{1em}
    \includegraphics[width=\columnwidth]{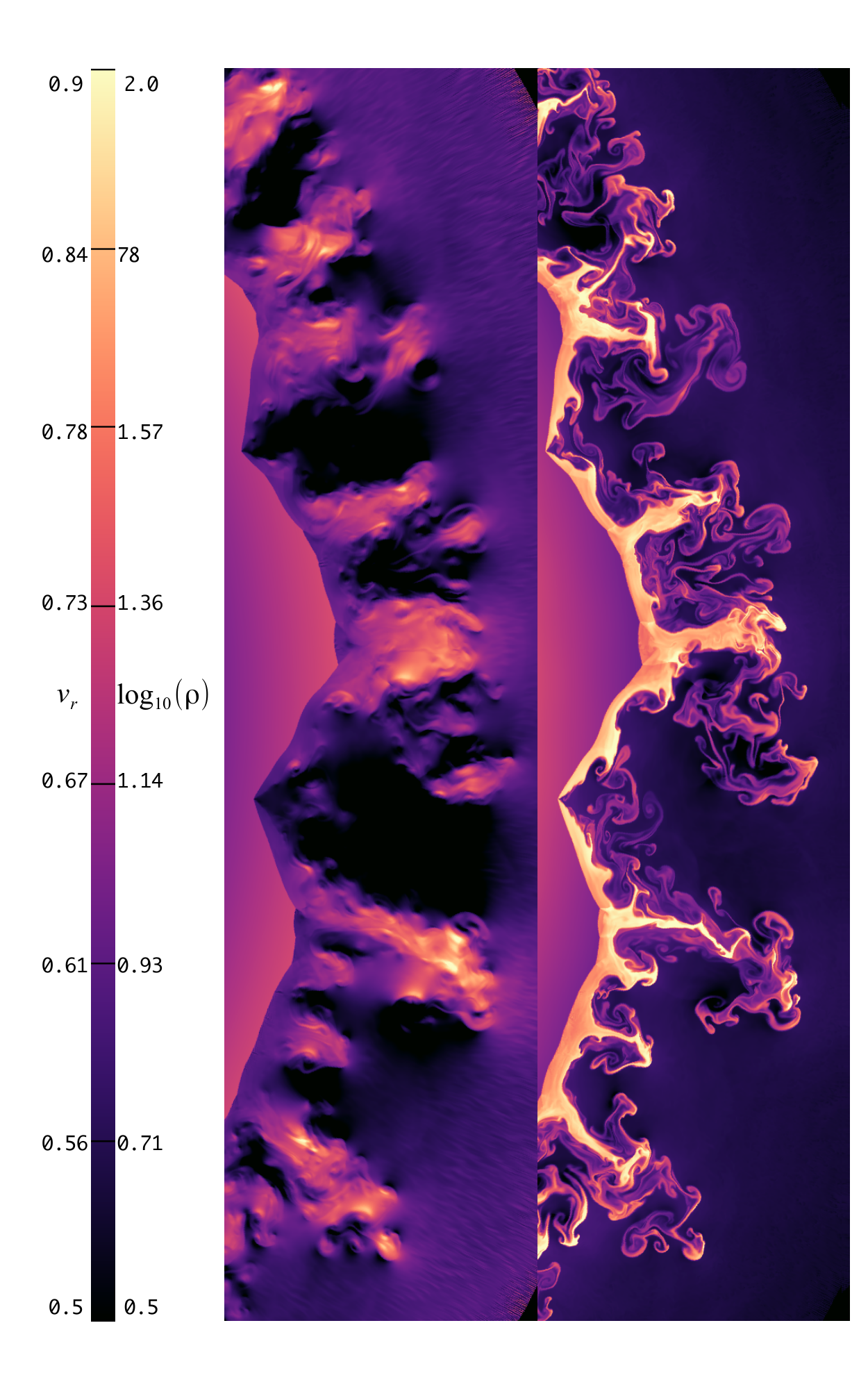}
    \caption{Rayleigh-Taylor fingers extending between the forward and reverse shock as the ejecta expands into the surrounding medium. Shown is the RTI arising from the Chevalier $s=2$, $n=7$ self-similar solution at a resolution of $8192\times8192$ ($r\times\theta$). Velocity is shown on the left and density on the right. The color bar displays values in code units.} 
    \label{fig:2dimg}
\end{figure}

\begin{figure*}
  \centering
  \includegraphics[width=\columnwidth]{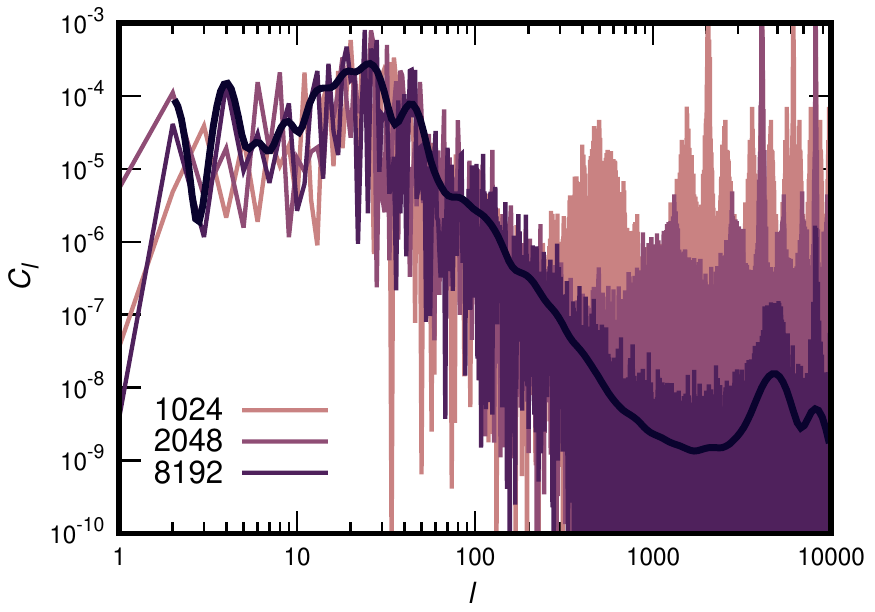}
  \includegraphics[width=\columnwidth]{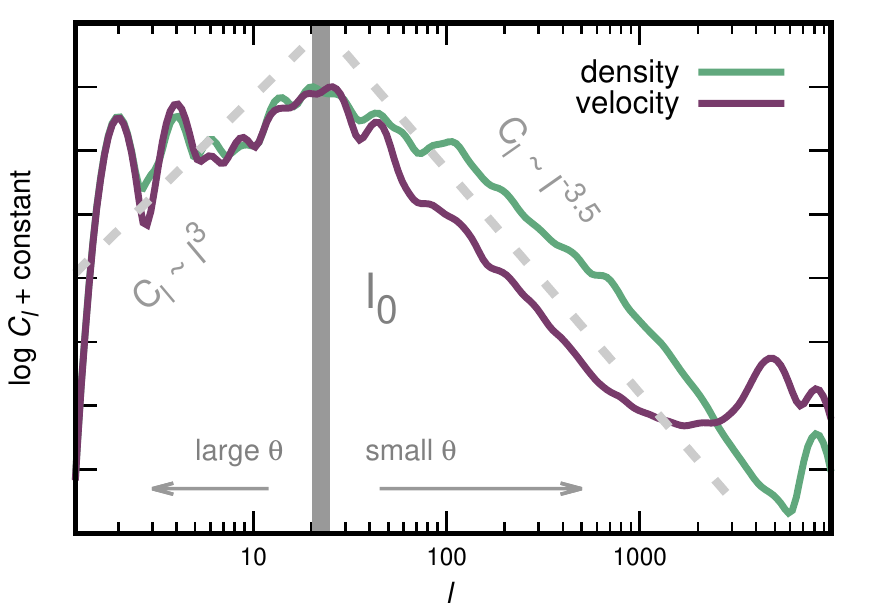} 
\caption{Power spectrum associated with the $s=2$, $n=7$ Chevalier solution. Left: Velocity power spectra at different computational resolutions. The dark line tracing the data is the data smoothed via a gaussian smoothing process used throughout the rest of this work for visualization purposes. Right: The smoothed power spectra of density (green) and velocity (purple) for the high resolution model. Dashed lines trace the slopes of the characteristic power laws and the solid vertical line denotes the dominant mode, $l_0$, at which they break. The secondary peak at high $l$ is caused by the grid scale.} 
\label{fig:s2n7resolutions}
\end{figure*}

\begin{figure}
  \centering
  \includegraphics[width=\columnwidth]{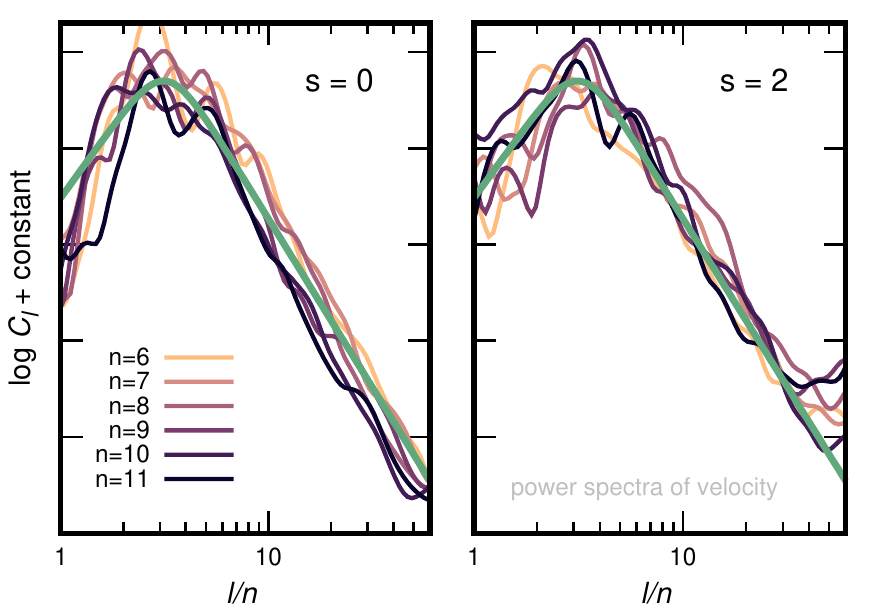} 
\caption{The power spectra associated with velocity for all models. When the x-axis is shifted by a factor of $n$ all of the power spectra align, making sense of the chaos. The green line denotes the best fit function (Equation~\ref{eq:psvelocity}). In this plot it becomes clear that $l_0 \sim 3n$.} 
\label{fig:data}
\end{figure}

\section{Results}
\label{sec:results}

Figure~\ref{fig:2dimg} shows a portion of our high resolution ($8192\times 8192$) model of the $s=2$, $n=7$ Chevalier solution. The left panel shows radial velocity and the right panel displays density. These plots are cropped to show the fine scale structure of the RTI turbulence, but the model is calculated for a full 2D axisymmetric ejecta.

The left panel of Figure~\ref{fig:s2n7resolutions} displays the power spectrum of the $s=2$, $n=7$ Chevalier solution calculated in three different resolutions, the highest resolution coming from the data in Figure~\ref{fig:2dimg}. The dark line tracing the high resolution data shows an example of a gaussian smoothing process we use for the remainder of the plots in order to visualize the character of these power spectra, but all analysis was performed on the un-smoothed data. The right panel shows the smoothed power laws corresponding to velocity (purple) and density (green) for the high resolution model. This figure also labels the different regions of the power spectra.

The behavior of power spectrum displays a broken power law, with a positive slope for small $l$, relating to large $\theta$ angular modes, and a negative slope for large $l$, small $\theta$ modes. These two slopes are broken at a value we call $l_0$, relating to the dominant angular mode, $\theta_0$, which separates large and small scale behavior ($\theta_0 = \pi/l_0$). A secondary peak is caused in the data by the grid scale, we see this peak move to higher $l$ as the resolution is increased. In order to properly characterize the slope of the small scale behavior, without contamination from the grid scale, we choose a minimum resolution of $2048\times2048$ ($r \times \theta$) for the remaining analysis. 

Figure~\ref{fig:data} shows the power spectrum of all of the hydrodynamic models. The two panels are the power spectra of the velocity where the left panels show all choices of n for the constant external density ($s=0$) case, and the right panels show the external wind ($s=2$) solutions. The y-axis is scaled to the maximum of $C_l$ and the x-axis is $l/n$. These are log-log plots so straight lines imply a power law in $l$. When plotted in this manner the power spectra sit on top of one another, exhibiting the same slopes and characteristic break (in terms of $l/n$). The green lines shows the best fit to our broken power law describing the broken power law. We find:

\begin{equation} \label{eq:psvelocity}
    C_{l} \propto \frac{1}{(l/3n)^{-3} + (l/3n)^{3.5} }
\end{equation}

To relate the power spectra to the physics of the ejecta we must note that physical quantities are represented not by the power, $C_l$, but by the amplitude, e.g. $\delta v_l \sim |a_{l,v}|=\sqrt{C_{l,v}}$. 

\subsection{The Dominant Angular Mode $\theta_0$}

\begin{figure}
  \centering
  \includegraphics[width=\columnwidth]{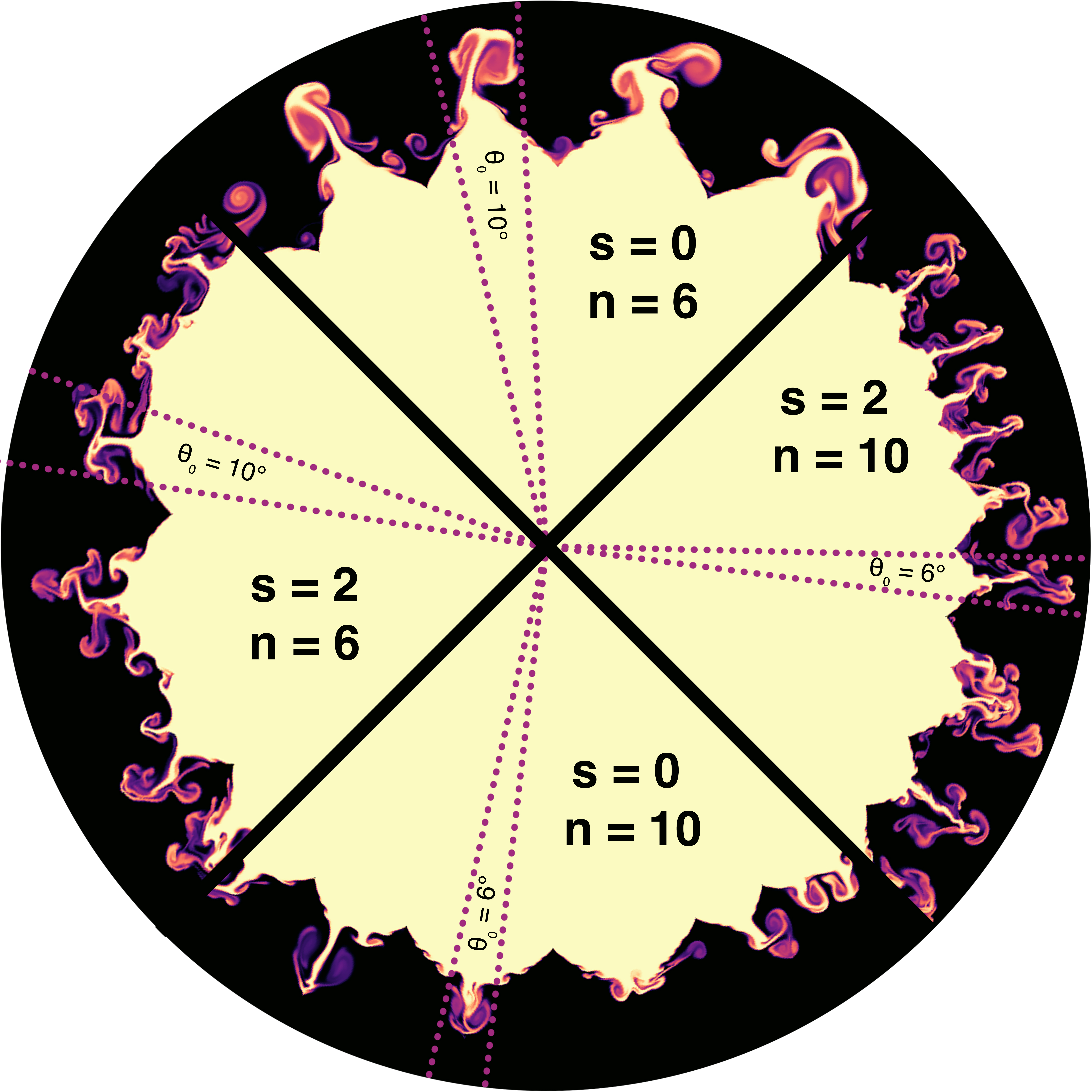}
\caption{Four example slices of the $2048 \times 2048$ resolution models. Plotted is a passive scalar where light yellow indicates fluid that originally belonged to the SN ejecta and dark purple indicates the surrounding material. Scales here are normalized by the radius of the shock. The dominant angular mode $\theta_0$ for each model is traced out by the dotted lines. This opening angle roughly traces the width of the Rayleigh-Taylor fingers.} 
\label{fig:2048thetas}
\end{figure}

The power spectra break at a value we have called the dominant angular mode $\theta_0=\pi/l_0$. This represents the highest amplitude angular scale in the SNR.

In Figure \ref{fig:data} we find the power spectrum peaks at a value 

\begin{equation}
    l_0 \approx 3n .
\end{equation}

In fact, one can compare by eye the value of $l_0$ to the to the character of the turbulence. Figure~\ref{fig:2048thetas} shows four examples of the $2048 \times 2048$ resolution models with the dominant angular mode traced out in each. $\theta_0$ is roughly the width of a Rayleigh-Taylor finger or similarly the angular spacing between fingers.

We interpret this empirically-derived scaling to mean that the dominant length scale is of order a density scale height (in the ejecta):

\begin{equation}
    h \equiv \left| \frac{\rho}{\nabla \rho} \right| = r/n.
\end{equation}

Noting that $l \approx \pi/\delta \theta$ corresponds to an angular scale, one can associate $l_0$ to a density scale height:

\begin{equation}
    h \sim \lambda = r \delta \theta \sim \pi r / l_0
\end{equation}
leading to the relationship

\begin{equation}
    l_0 \approx \pi n.
\end{equation}

We find a relationship $l_0 \approx 3 n$ empirically, but the coefficient may change in 3D.  In fact, 3D studies by \cite{Warren2013} suggest that the coefficient is larger, around $l_0 \approx 10 n$ in 3D, but a systematic parameter survey is necessary to confirm this.

The value of $l_0$ is perhaps the most powerful diagnostic in the power spectrum. The $s=0$ case and $s=2$ case follow the same relationship for  $l_0$, suggesting that the break in the power spectrum only depends on $n$. This is a direct measurement that can be made from the SNR to reveal the density gradient of the outer layers of SN ejecta. This density gradient can vary in theoretical explosion models in both core collapse SN models and for different SN~Ia progenitor explosion mechanisms. 

\cite{Warren2013} present 3D models for a SNR arising from an exponential ejecta profile (typical of Type Ia SNe). The power spectra of their models show a value of $l_0$ that changes with time, going from small scales to larger scales as the remnant evolved. This is consistent with our explanation of the density scale height determining $l_0$. In the Chevalier models the scale height does not evolve with time once self-similarity is achieved. For an exponential profile, however, as the reverse shock overtakes more of the ejecta, the density scale height becomes larger with time (relative to the radius of the reverse shock, $R_{RS}$).  For an ejecta model with

\begin{equation}
    \rho \propto e^{-v/v_e}
\end{equation}
with $v = r/t$ and $v_e$ is the characteristic velocity of the ejecta, the density scale height is given by

\begin{equation}
    h = v_e t
\end{equation}

Thus, one would expect an ``effective" value of $n$ given by

\begin{equation}
    n_{\rm eff} = r/h = \frac{R_{RS}}{v_e t} 
\end{equation}
where the relevant radius can be taken as the radial position of the reverse shock.  

\cite{Warren2013} reported all of this information for their models: the velocity $v_e = 0.289$ (all quantities here in code units).  At time $t = 0.12$, the reverse shock had progressed to $R_{\rm RS} = 0.249$ with $l_0 \approx 61$.  At $t = 0.75$, $R_{\rm RS} = 0.716$ and $l_0 \approx 42$.  At $t = 2.0$, $R_{\rm RS} = 1.005$ and $l_0 \approx 25$ (values of $l_0$ estimated by eye from their Figure 3).  Thus, $l_0/n_{\rm eff}$ at these three times is equal to $8.5$, $12.7$, and $14.4$ respectively.  These are reasonably consistent with a scaling of $l_0 \approx 10 n$, a peak at somewhat smaller angular scales than we find in 2D.

\subsection{Power-law Slopes}

On either side of $l_0$, the power spectrum is very steeply broken.  The sharp peak at $l_0$ suggests that we are not seeing evidence of a turbulent cascade, but of the driving and saturation of the instability at each given scale in isolation.  In this case, it is possible that 2D and 3D calculations might be expected to exhibit similar scalings, so long as they are normalized in a consistent way.  Systematic 3D studies are needed to determine this.

In 3D, Kolmogorov turbulence is determined by the rate of transfer of energy from scale to scale.  For uniform 3D turbulence, the power is independent of the scale \citep{Kolmogorov1941}:

\begin{equation}
    \frac{\delta v^2}{\tau} = \text{constant}
\end{equation}

where $\tau=\lambda/\delta v$ is an eddy turn-over time, and $\lambda \sim r/l$ is the eddy size. So proportionally:

\begin{equation}
    |\delta v| \propto \lambda^{-1/3} \propto (r/l)^{-1/3} \qquad \text{or} \qquad C_l \propto |\delta v|^2/l \sim l^{-5/3}
\end{equation}

In 2-dimensions this becomes slightly different with the added conservation of vorticity, which gives a cascade of constant $\tau = \lambda/\delta v= \text{constant}$:

\begin{equation}
    |\delta v| \propto \lambda \propto l^{-1} \qquad \text{or} \qquad C_l \propto l^{-3}
\end{equation}

Our small-scale anisotropies exhibit a steeper scaling than either of these slopes, with $C_l \propto l^{-3.5}$, and 3D studies by \cite{Warren2013} also see a very steep scaling ($C_l \propto l^{-3.9})$, inconsistent with Kolmogorov turbulence.  Thus, we might not be witnessing a turbulent cascade, but the growth and saturation of RTI at each scale independently.

So why do we see no turbulent cascade?  Fluctuations are developing, with non-zero kinetic energy; one would ordinarily expect that energy to cascade.  It may be the case that eddies do not have time to transfer their energy to smaller scales (or to larger ones); if the eddy turn-over time is longer than an expansion time or an RTI growth time, then the turbulent cascade might not be efficient enough to move the developed kinetic energy from one scale to another.
\\
\subsection{Large Scale Behavior}

The large scale, small $l$ behavior is not as obviously described by a simple power law, however it is clear that the power grows steeply with $l$ until the value of $l_0$ is reached. As such we employ a power law fit and discuss the implications of this growth here.

The power-law for large scales is quite steep, $C_l \propto l^3$ for $l<l_0$.  Therefore, RTI has very little effect at these large scales. \cite{Chevalier1992} found that RTI does not grow for $l \lesssim 7$ (in the $n=7$, $s=2$ case).  For the regime between this cutoff and $l_0$, RTI can grow but it appears to be suppressed by the density gradient in the ejecta (which cannot be detected by small-scale modes). What this means is that the large scale structure is where information about the underlying SN can leave significant signatures on the remnant without interference from RTI mixing. For the purely spherical case of the Chevalier outflow we see slope of $+3$, but anisotropies in the explosion could leave a detectable signature in this regime. For example \cite{Lopez2011} and \cite{Lopez2018} show that power in the $l=2$ and $l=3$ modes can be used to distinguish between a SNR arising from a thermonuclear explosion and one resulting from a core collapse supernova. Any additional power scales in this large $\theta$ regime are what we should examine while trying to understand the fundamental nature of these explosions.  

\cite{Ferrand2019,Ferrand2021,Ferrand2022} present a series of SNe Ia explosion models expanded until they reach the remnant phase, and calculate power spectra corresponding to the forward-shock, reverse-shock and contact discontinuity. For the case of a 3D spherically symmetric model the primary of character of the power spectra are consistent with those presented here. However, the cases where the initial models are not spherically symmetric result in bimodal power spectra. The second peak (around $l\sim40$) is consistent with RTI (and with our results here) and the first peak is entirely generated from initial large-scale asymmetries in the explosion. More systematic studies of how SNe~Ia models impart power on these larger angular scales could lead to another method for distinguishing between Ia progenitor theories with SNRs.

As \cite{Warren2013} showed, at sufficiently late times, the density gradient becomes large enough relative to the reverse shock radius that RTI can grow to much larger scales.  Therefore, it might be theoretically possible for RTI to impose large-scale asymmetries in a SNR, but only as the reverse shock is sweeping through most of the mass in the ejecta.  Additionally, RTI in a pulsar wind nebula such as the Crab might be able to impose asymmetries on larger scales than a typical SNR, because the density gradient is not as steep in this case.  3D calculations of RTI in pulsar wind nebulae are needed in order to confirm or disprove this hypothesis.

\newpage
\section{External Clumps}
\label{sec:clumpy}

\begin{figure}
  \centering
  \includegraphics[width=\columnwidth]{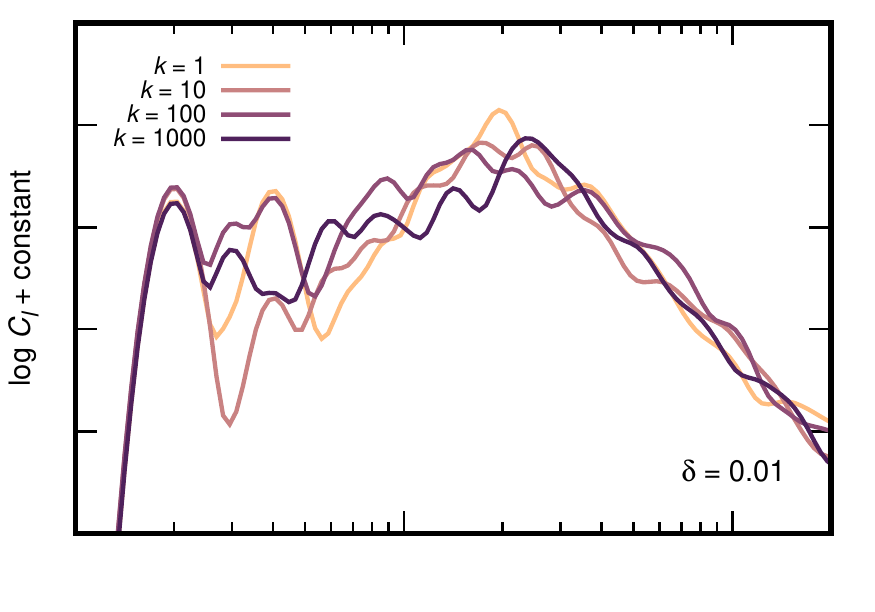} \vspace{-4em} \\
  \includegraphics[width=\columnwidth]{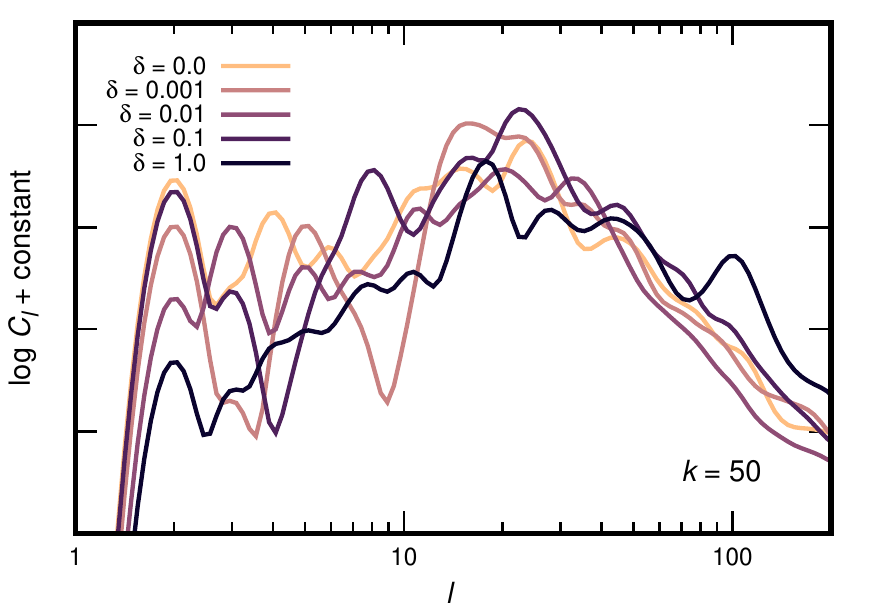}
\caption{Effect on the power spectra of varying perturbations in the surrounding medium medium of the n=7 s=2 case according to Equation~\ref{eq:clumpy}. Each y-axis tilt represents one order in $C_l$. The clumpy medium does not change the character of the SNR power spectrum.} 
\label{fig:clumpy}
\end{figure}

There has been much discussion in literature debating whether clumping observed in SNRs is inherent to the SN or arises from a non-uniform, or ``clumpy" external medium (see for e.g. \cite{Celli2019,Sano2020,Sano2021,Tanaka2020,Fujita2022}). Here we examine the effect of a CSM with density variations on the power spectra of the ejecta.

For the $n = 7$, $s = 2$ case we place density perturbations into the outer medium with the function:

\begin{equation} \label{eq:clumpy}
    \Delta \rho_{pert} = \delta \sin(k \log(r))\sin(k\theta)
\end{equation} 

Figure~\ref{fig:clumpy} shows the results of this exercise. The top panel displays the power spectra keeping the magnitude of these perturbations ($\delta$) constant while varying the wavelength ($k$). When varying $k$ no change in the character of the power spectrum is detected; both power law slopes and the break in the spectrum remain the consistent. The bottom panel shows a constant choice for $k$ but this time varying the amplitude of the perturbations. Here the spectra are noisier but the characteristic slopes and dominant mode $l_0$ still remain unchanged. 

Only when order-unity fluctuations are introduced, so that $\rho_{\rm max}/\rho_{\rm min} \approx 7$, at a very narrow (coherent) wavelength of $k = 50$, can the power spectrum be affected.  But this is an extreme example; generally speaking, as the forward shock overtakes clumps in the ejecta, it is stable to perturbations and flattens them out, so that these clumps merely introduce a small seed to the growth of the instability.  The final state of the fluid is, generally speaking, unaffected by external clumps.  This result was also found by \cite{Chevalier1992} and by \cite{Warren2013}.

It is worth making the distinction between external clumps, tested here, and {\em internal} clumps, or anisotropies in the ejecta.  It was shown by \cite{Orlando2012} that {\em internal} clumps can significantly affect the final state of the fluid.  This is another way of saying that asymmetries in the initial explosion can be preserved, especially if they exist on sufficiently large angular scales.

\section{Summary}
\label{sec:discussion}

Power spectral analysis is a powerful tool with which to understand SNRs. In this work we have presented template SNR power spectra and identified the characteristics with which we can learn about the physics of the explosion. The important diagnostics are as follows:

\begin{itemize}
    \item The power spectra take the form of a broken power law where the break, $l_0$, represents the largest angular scale ($\theta_0=\pi/l_0$) that can grow via RTI.
    \item This mode, $l_0$ is determined by the density scale height and therefore is a direct diagnostic of the density profile of the outer SN ejecta layers. Empirically we find $l_0\sim3n$ where $n$ is the power law describing the SN ejecta profile.
    \item The power spectrum for $l>l_0$ show a power law slope of $C_l\propto l^{-3.5}$ which is steeper than expected for a Kolmogorov turbulent cascade. Therefore small scale behavior may instead be driven by the saturation of RTI at these scales.
    \item The $l<l_0$ modes are where the SN ejecta can imprint itself upon the power spectra. Additional power at these larger angular scales in remnant data indicates anisotropies endemic to the SN itself.
    \item The power spectra of these remnants are stable to varying conditions of the gas external to the SN. We show that clumpiness in the surrounding medium will not impose additional power in the the SNR characteristics.
\end{itemize}

Work by \cite{Warren2013} providing a 3D example of an exponential density profile running into a surrounding medium indicates that the slopes of the power spectra are fairly consistent between our 2D models and a 3D simulation. However the precise dependence of $l_0$ seems to shift to slightly smaller scales; $l_0\sim10n$ in 3D compared to our $l_0\sim3n$ in 2D. A systematic 3D study is required to investigate this difference.

The models and power spectra presented in this work are most directly relevant for young SNRs, specifically those for which the reverse shock is still present and has not yet propagated through the full ejecta. This method, however, can and should be applied to a larger variety of SNR models, for example older remnants where the reverse shock has propagated through the full ejecta, jet powered SNe and ejecta with an energy source to model pulsar wind nebulae. Power spectral analysis should also be performed on observed SNRs. By comparing what we learn from the power spectra of models to the power spectra of observations we can make quantitative inferences about the SN explosions from the morphology of their remnants. 

\section*{Acknowledgements}
We would like to thank the anonymous referee for the helpful review. Numerical calculations were performed on the Stampede2 supercomputer under allocations TG-PHY210027 \& TG-PHY210035 provided by the Extreme Science and Engineering Discovery Environment
(XSEDE), which is supported by National Science Foundation grant number ACI-1548562 \citep{Towns2014}.  P.~D.\ acknowledges NSF support from grant AAG-2206299, and NASA support from grant 21-FERMI21-0034. D.~M.\ acknowledges NSF support from grants PHY-1914448, PHY-2209451, AST-2037297, and AST-2206532. This research was supported in part by the National Science Foundation under Grant No. NSF PHY-1748958.

\software{JET \citep{2013ApJ...775...87D}}

\bibliographystyle{aasjournal}
\bibliography{main.bib}

\end{document}